\documentclass{article}
\usepackage{graphics}
\usepackage{amsmath}
 \usepackage{epsfig}
\usepackage{amssymb}
\usepackage{amsfonts}
 \usepackage{latexsym}
\begin{document}

\parskip=-0.cm

\newtheorem{Proposition}{Proposition}
\newtheorem{Comment}{Proposition}
  \newtheorem{Remark}[Proposition]{Remark}
  \newtheorem{Corollary}[Proposition]{Corollary}
  \newtheorem{Lemma}[Proposition]{Lemma}
    \newtheorem{Theorem}[Proposition]{Theorem}
  \newtheorem{Note}[Proposition]{Note}
\newtheorem{Definition}{Definition}
\def\bfy{\mathbf{y}}
\def\bfz{\mathbf{z}}
\def\bfC{\mathbf{C}}
\def\Nsm{\hbox{\small I\hskip -2pt N}}
\def\rdb{\hbox{ I\hskip -2pt R}}
\def\cdb{\hbox{\it l\hskip -5.5pt C\/}}
\def\ndd{\hbox{\it I\hskip -2pt N}}
\def\zdd{\hbox{\sf Z\hskip -4pt Z}}
\def\e{\epsilon}
\def\ei{\epsilon^{-1}}
\def\a{\alpha}
\def\l{\lambda}
\def\tl{\tilde\lambda}
\def\ct{{\mbox{const}}}
\def\d{\delta}
\def\cf{{\cal F}}
    \def\z{\noindent}  
 \def\Box{{\hfill\hbox{\enspace${\sqre}$}} \smallskip}
    \def\sqr#1#2{{\vcenter{\vbox{\hrule height .#2pt
                             \hbox{\vrule width .#2pt height#1pt \kern#1pt
                                   \vrule width .#2pt}
                             \hrule height .#2pt}}}}
 \def\sqre{\mathchoice\sqr54\sqr54\sqr{4.1}3\sqr{3.5}3}     
     \def\erm{\mathrm{e}}
    \def\irm{\mathrm{i}}
    \def\drm{\mathrm{d}}
 \def\bchi{\mbox{\raisebox{.4ex}{\begin{Large}$\chi$\end{Large}}}}
     \def\CC{\mathbb{C}}
    \def\DD{\mathbb{D}}
    \def\NN{\mathbb{N}}
    \def\QQ{\mathbb{Q}}
    \def\RR{\mathbb{R}}
    \def\ZZ{\mathbb{Z}}

\title{Nonperturbative analysis of a model quantum system under time
periodic forcing}

\gdef\shorttitle{Periodically forced quantum system: nonperturbative analysis}
\author{Ovidiu Costin$^*$, Rodica D. Costin$^*$, Joel. L. Lebowitz$^*$
  and Alexander Rokhlenko\footnote{Department of Mathematics, Hill Center,
  Rutgers University, New Brunswick, NJ 08903, USA.  e-mail:
  costin\symbol{64}math.rutgers.edu, rcostin\symbol{64}math.rutgers.edu,
  lebowitz\symbol{64}math.rutgers.edu,
  rokhlenk\symbol{64}math.rutgers.edu.  }}

\maketitle

\vskip 0.5cm 

\begin{abstract}
  We analyze the time evolution of a one-dimensional quantum system with
  an attractive delta function potential whose strength is subjected to
  a time periodic (zero mean) parametric variation $\eta(t)$.  We show
  that for generic $\eta(t)$, which includes the sum of any finite
  number of harmonics, the system, started in a bound state will get
  fully ionized as $t\rightarrow\infty$ irrespective of the magnitude or
  frequency of $\eta(t)$. For the case $\eta(t)=r\sin(\omega t)$ we find
  an explicit representation of the probability of ionization.  There
  are however exceptional, very non-generic $\eta(t)$, that do not lead
  to full ionization. These include rather simple explicit periodic
  $\eta(t)$ for which the system evolves to a nontrivial localized
  stationary state related to eigenfunctions of the Floquet operator.
\end{abstract}

\smallskip
\centerline{{\Large  Analyse non-perturbative d'un systeme quantique mod\`ele
avec force exterieure}}\centerline{\Large periodique}

\smallskip

\centerline{\bf R\'esum\'e}
\begin{description}
\item{\ \ \ \ \ \ Nous analysons l'\'evolution dans le temps d'un
    syst\`eme unidimensionel avec un potentiel attractif de type
    fonction delta soumis a une variation p\'eriodique de moyenne nulle,
    $\eta(t)$. Nous d\'emontrons que pour $\eta$ g\'en\'erique (en
    particulier pour une somme finie d'oscillations harmoniques) le
    syst\`eme qui est d'abord dans un \'etat li\'e vat \^etre
    compl\`etement ionis\'e pour $t\rightarrow\infty$. Des fonctions
    $\eta(t)$ tr\`es nong\'en\'eriques, toutefois explicites,
    existent pour lesquelles le syst\`eme \'evolue vers un \'etat
    localis\'e non-trivial, li\'e aux fonctions propres de l'op\'erateur
    de Floquet associ\'e.}
\end{description}

\section{Version fran\c caise abr\'eg\'ee}

Nous \'etudions rigoureusement le comportement pour $t\rightarrow\infty$
d'un
syst\`eme quantique  unidimensionnel simple, avec potentiel attractif de
type delta,
 soumis \`a une variation param\'etrique
p\'eriodique. Dans des unit\'es convenables, le Hamiltonien est de la
forme

$$H(t)=H_0-2\, \eta(t)\delta(x)= \frac{\mathrm{d}^2}{\mathrm{d}x^2}-
2\,\delta
  (x)-2\, \eta(t)\delta(x)
$$ 

\z o\`u $H_0$ a un seul \'etat li\'e $u_b=e^{-|x|}$ d'\'energie
$\-\omega_0=-1$ et un spectre continu sur l'axe r\'eel positif, avec
fonctions propres g\'en\'eralis\'ees, voir eq.  (\ref{ef}).

On peut d\'evelopper la solution de l'\'equation de Schr\"odinger
$\psi(x,t)$ par rapport aux fonctions propres de $H_0$ (\ref{eq:(2)})
avec conditions initielles $\theta (0)=\theta_0,\ \Theta
(k,0)=\Theta_0(k)$ normalis\'ees convenablement, eq. (\ref{eq:e4}). Alors,
la probabilit\'e de survie de l'\'etat li\'e est $P(t)= |\theta(t)|^2$,
alors que $|\Theta(k,t)|^2 dk$ donne la ``fraction de particules
\'eject\'ees'' avec (quasi-) impulsion dans l'intervalle $dk$. En
prennant la fonction $Y$ donn\'ee par (\ref{eq:eqY}) on obtient les
\'equations eq. (\ref{eq:(3)}) et $Y$ satisfait une \'equation int\'egrale,
(\ref{eq:(5)}). Notre m\'ethode d'analyse utilise les propri\'et\'es
analytiques de la transformation de Laplace $y(p)$ de $Y(t)$ pour
d\'eterminer les propri\'et\'es asymptotiques de $Y(t)$ par rapport \`a $t$.

\section{Cas o\`u $\eta(t)$ est harmonique}

\z {\bf Th\'eor\`eme 1 \cite{[19]}}. {\em Si
$\eta(t)=r\sin\omega t$ ,
 la probabilit\'e de survie $|\theta(t)|^2$ tend vers
z\'ero quand $t\rightarrow\infty$, pour tous $\omega>0$ et $r\ne 0$.}

\z {\bf Remarques}. ({\bf 1}) On obtient une formule exacte
(\ref{eq:intform}) pour $\theta(t)$ o\`u $F_\omega$ est p\'eriodique de
p\'eriode $2\pi\omega^{-1}$ et les fonctions $h_m$ satisfont
(\ref{eq:defh}).  Pas trop pr\`es des r\'esonances, si
$|\omega-n^{-1}|>O(r^{2-\delta})$ pour tout entier positif $n$,
$|F_{\omega}(t)|=1\pm O(r^2)$ et les coefficients de Fourier de $F$
d\'ecroissent plus vite que $r^{|2m|} |m|^{-|m|/2}$.  Aussi, la somme en
(\ref{eq:intform}) est plus petite que $O(r^2 t^{-3/2})$ pour $t$ grand,
et les $h_m$ d\'ecroissent avec $m$ plus vite que $r^{|m|}$.

({\bf 2}) On voit gr\^ace \`a (\ref{eq:intform}) que pour des temps d'ordre
$1/\Gamma$
o\`u
$\Gamma=2\Re(\gamma)$, la probabilit\'e de survie pour $\omega$ pas trop
pr\`es
d'une r\'esonance d\'ecroit comme $\exp(-\Gamma t)$, et que 
le comportement asymptotique en $t$ est  $|\theta(t)|^2=O(t^{-3})$ avec
beaucoup d'oscillations.

({\bf 3}) Quand $r$ est plus grand, le comportement
polynomial-oscillatoire commence plus t\^ot et la probabilit\'e de
survie est plus {\em grande}. Ce ph\'enom\`ene est parfois appel\'e
stabilisation atomique.

({\bf 4}) En utilisant une fraction continue convergente, on peut
calculer $\Gamma$ pour $\omega$ and $r$ arbitraires.  Pour $r$ petit, si
$n$ est la partie enti\`ere de $\omega^{-1}+1$ et si
$\omega^{-1}\notin\NN$, alors, pour $T>0$ ($t=r^{-2n}T$), $\Gamma$ est
donn\'e par (\ref{limit}).

({\bf 5}) Le comportement de $\Gamma$ est diff\'erent aux r\'esonances
$\omega^{-1}\in\NN$. Par exemple si $\omega-1=r^2/\sqrt{2}$ on trouve
la formule (\ref{reson}).

\subsection{Cas  p\'eriodique g\'en\'eral}
On \'ecrit $\eta$ sous la forme (\ref{defeta}) et nos
hypoth\`eses sur les $C_j$ sont 
(a) $0\not\equiv\eta\in L^{\infty}(\mathbb{T})$, (b) $C_0=0$ et (c)
$C_{-j}=\overline{C_j}$. Consid\'erons aussi l'hypoth\`ese de
g\'en\'ericit\'e {\bf (g)} suivante:  on d\'efinit la translation 
a droite ${T}(C_1,C_2,...,C_n,...)=(C_2,C_3,...,C_{n+1},...)$. 
Alors, ${\bf  C}\in l_2(\NN)$ est {\em g\'en\'erique par rapport a $T$}
si l'espace de Hilbert engendr\'e par toutes les translations de
${\bf C}$ contient le
vecteur $e_1=(1,0,0...,)$, cf. (\ref{eq:cyclic}). Un cas important 
est donn\'e par les polyn\^omes trigonom\'etriques. Un exemple qui 
ne satisfait pas \`a l'hypoth\`ese (g) est (\ref{eq:counterex}) pour
$\lambda\in (0,1)$, pour lequel $C_n=-r\lambda^n$ for $n\geq
1$.

\subsection{Resultats dans le c\`as p\'eriodique}

{\bf Th\'eor\`eme 2 \cite{[21]}}. {\em 
 Sous les hypoth\`eses (a), (b), (c) et (g), la probabilit\'e de survie
  $P(t)$ 
de l'\'etat li\'e $u_b$, 
  $|\theta(t)|^2$ tend vers z\'ero quand $t\rightarrow\infty$.}

\z {\bf Th\'eor\`eme 3 \cite{[21]}}. {\em 
  Pour $\psi_0(x)=u_b(x)$  il existe des valeurs de
  $\lambda$, $\omega$ et $r$ en (\ref{eq:counterex}), pour
lesquelles
  $|\theta(t)|\not\rightarrow 0$ quand
  $t\rightarrow\infty$.}

\z {\bf Remarque}. Le Th\'eor\`eme~3 peut \^etre \'etendu pour
d\'emontrer que pour $r$ et $\omega$ donn\'es en (\ref{eq:counterex}) il
existe un ensemble infini de $\lambda$, avec un point d'accumulation en
$1$, pour lequel $\theta(t)\not\rightarrow 0$.

\section{English version}

\subsection{The problem}
\label{intro}

We are interested in the  nature of
the solutions of the Schr{\"o}dinger equation 

\begin{equation}
  \label{eq:Sc1}
  i\hbar\partial_t\psi=[H_0+H_1(t)]\psi
\end{equation}
Here $\psi$ is the wavefunction of the system, belonging to some Hilbert
space ${\cal H}$, $H_0$ and $H_1$ are Hermitian operators and
equation (\ref{eq:Sc1}) is to be solved subject to some initial condition $\psi_0$.
$H_0$ has both a discrete and a continuous spectrum corresponding
respectively to spatially localized (bound) and scattering (free) states
in $\RR^d$. Starting at time zero with the system in a bound state and
then ``switching on'' at $t=0$ an external potential $H_1(t)$, we want
to know the ``probability of survival'', $P(t)$, of the bound states, at
times $t>0$; $1-P(t)$ is the probability of ionization
\cite{[2]}--\cite{[8]}.

 When $\omega$ is sufficiently large for ``one photon'' ionization to
take place, i.e., when $\hbar\omega>-E_0$, $E_0$ the energy of the bound
(e.g.\ ground) state of $H_0$ and $r$ is ``small enough'' for $H_1$ to
be treated as a perturbation of $H_0$ then  the long time behavior of
$P(t)$ is asserted in the physics literature to be
given by  $P(t)\sim \exp[-\Gamma_F t]$. The rate constant $\Gamma_F$
is computed from first order perturbation theory according to Fermi's
golden rule.  It is proportional to the square of the matrix element
between the bound and free states, multiplied by the appropriate density
of continuum states in the vicinity of the final state which will have
energy $\hbar\omega+E_0$ \cite{[6],[7],[8]}.

The results described here show that the phenomenon of ionization by
periodic fields is very complex indeed once one goes beyond the
perturbative regime.
\section{Our model}
\label{model}

We consider a very simple quantum system where we can analyze rigorously
many of the phenomena expected to occur in more realistic systems
described by (\ref{eq:Sc1}).  This is a one dimensional system with an
attractive delta function potential.  The unperturbed Hamiltonian $H_0$
has, in suitable units, the form

\begin{equation}
  \label{eq:(1)}
  H_0=-{\frac{\mathrm{d}^2}{\mathrm{d}x^2}}-
2\,\delta
  (x),\ \ -\infty<x<\infty.
\end{equation}
 $H_0$ has a single bound state $ u_b(x)=e^{-|x|}$ with
energy $ -\omega_0=-1$. It also has continuous uniform spectrum on the
positive real line, with generalized eigenfunctions

\begin{equation}
  \label{ef}
u(k,x)=\frac{1}{\sqrt{2\pi}}\left
(e^{ikx}-\frac{1}{1+i|k|}e^{i|kx|} \right ), \ \ -\infty<k<\infty  
\end{equation}

\z and energies $k^2$.  

Beginning at  $t=0$, we apply a parametric perturbing
potential, i.e. for $t>0$ we have

\begin{equation}
  \label{eq:timedep}
  H(t)=H_0 -2\, \eta(t)\delta(x)
\end{equation}
\z and  solve the time  dependent Schr{\"o}dinger
equation (\ref{eq:Sc1})
 for $\psi(x,t)$, with $\psi(x,0) = \psi_0(x)$.  Expanding $\psi$ in
eigenstates of $H_0$ we write
\begin{equation}
  \label{eq:(2)}
 \psi (x,t)=\theta (t)u_b(x)e^{it}+\int_{-\infty}^
{\infty}\Theta (k,t)u(k,x)e^{-i k^2 t}dk \ \ (t\geq 0)
\end{equation}
with initial values $\theta (0)=\theta_0,\ \Theta (k,0)=\Theta_0(k)$  suitably normalized, 

\begin{equation}
  \label{eq:e4}
\langle \psi_0, \psi_0 \rangle =  |\theta_0|^2+\int_{-\infty}^\infty|\Theta_0(k)|^2dk=1
\end{equation}

We then have that the survival probability of the bound state is  $P(t) =
|\theta(t)|^2$, while  $|\Theta(k,t)|^2 dk$ gives the  
``fraction of ejected particles'' with 
(quasi-) momentum in the interval $dk$.

This problem can be reduced to the solution of an integral equation in a
single variable \cite{[19]}--\cite{[23]}.  Setting

\begin{equation}
  \label{eq:eqY}
  Y(t)=\psi(x=0,t)\eta(t)e^{it}
\end{equation}

\z we have

\begin{eqnarray}
  \label{eq:(3)}
  &\theta (t)=\theta_0+2i\int_0^t Y(s) ds \ , \\
  &\Theta(k,t)= \Theta_0(k)+2|k|/\big[\sqrt{2\pi} (1-i|k|)\big]\int_0^t Y(s)
e^{i(1+k^2)s} ds \ .
\end{eqnarray}

\z  $Y(t)$ satisfies the integral equation

\begin{equation}
  \label{eq:(5)}
  Y(t)=\eta(t)\left
\{I(t)+\int_0^t [2i+M(t-t')]Y(t') dt'\right \}=\eta(t)\Big(I(t)+(2i+M)*Y\Big)
\end{equation}

\z where the inhomogeneous term is

$$I(t)=\theta_0+\frac{i}{\sqrt{2\pi}}\int_0^{\infty}\frac{\Theta_0(k)+\Theta_0(-k)}{1+ik}e^{-i(k^2+1)t}dk,$$

\z and

$$
M(s)=\frac{2i}{\pi}\int_0^\infty \frac{u^2e^{-is(1+u^2)}}{1+u^2}du=
\frac{1+i}{2\sqrt{2}\pi}\int_s^\infty\frac{e^{-iu}}{u^{3/2}} du
$$ 

\z with

$$f*g=\int_0^tf(s)g(t-s)ds$$

{\bf Outline of the technical strategy}.  The method of analysis in
\cite{[19],[21]}, relies on the properties of the Laplace transform of
$Y$, $y(p)=\mathcal{L}Y(p)=\int_0^{\infty}e^{-pt}Y(t)dt$.

 We show that (\ref{eq:(5)}) has a unique solution in suitable
norms. This solution is Laplace transformable and the Laplace transform
$y$ satisfies a linear functional equation.  The solution of the
functional equation is unique in the right half plane provided it
satisfies the additional property that $y(p_0+is)$ is square integrable
in $s$ for any $p_0>0$. We use the functional equation to determine the
analytic properties of $y(p)$.

This is done using (appropriately refined versions of) the Fredholm
alternative. After some transformations, the functional equation reduces
to a linear inhomogeneous recurrence equation in $l_2$, involving a
compact operator depending parametrically on $p$.  The dependence is
analytic except for a finite set of poles and square-root branch-points
on the imaginary axis and we show that the associated homogeneous
equation has no nontrivial solution. We then show that the poles in the
coefficients do not create poles of $y$, while the branch points are
inherited by $y$. The decay of $y(p)$ when $|\Im(p)|\rightarrow\infty$,
and the degree of regularity on the imaginary axis give us the needed
information about the decay of $Y(t)$ for large $t$.

\subsection{Case when $\eta(t)$ is harmonic}

\begin{Theorem}[\cite{[19]}]\label{T1}
 When $\eta(t)=r\sin\omega t$ 
 the survival probability $|\theta(t)|^2$ tends to
zero as $t\rightarrow\infty$, for all $\omega>0$ and $r\ne 0$.  
\end{Theorem}

\z {\bf Remarks}. ({\bf 1}) The detailed behavior of the system as a function of $t$,
$\omega$, and $r$ is obtained from the singularities of $y(p)$ in the
complex $p$-plane. We summarize them {\em for small $r$}; below
$\delta>0$. For definiteness we assume in the following that $r>0$.

At $p=\{i n\omega-i: n\in\ZZ\}$, $y$ has square root branch points and
$y$ is analytic in the right half plane and also in an open neighborhood
${\mathcal{N}}$ of the imaginary axis with cuts through the branch
points. As $|\Im(p)|\rightarrow\infty$ in ${\mathcal{N}}$ we have
$|y(p)|=O(r \omega |p|^{-2})$.  If $|\omega-\frac{1}{n}|> {\rm
const}_nO(r^{2-\delta}) ,\,n\in\ZZ^+$, then for small $ r $ the function
$y$ has a unique pole $p_m=p_0+im\omega$ in each of the strips $
-m\omega>\Im(p)+1\pm O(r^{2-\delta}) >-m\omega-\omega,\,\ m\in\ZZ$.
$\Re(p_m)$ is strictly independent of $m$ and gives the exponential
decay of $\theta$. 
Laplace transform techniques show that

\begin{eqnarray}
  \label{eq:intform}
 \theta(t)=
e^{-\gamma(r;\omega) t}F_\omega(t)+\sum_{m=-\infty}^\infty
e^{(mi\omega-i)t}h_m(t)&
\end{eqnarray}

\vskip -0.2cm 

\z where $F_\omega$ is periodic of period $2\pi\omega^{-1}$
and 

\begin{equation}
  \label{eq:defh}
h_m(t)\sim \sum_{j=0}^{\infty}c_{m,j}t^{-3/2-j}\ \
\mbox{as }t\rightarrow\infty, \ \arg(t)\in
\Big(-\frac{\pi}{2}-0,\frac{\pi}{2}+0\Big)  
\end{equation}

\z {\em Not too close to resonances}, i.e.\ when
$|\omega-n^{-1}|>O(r^{2-\delta})$, $\delta>0$, for all integer $n$,
$|F_{\omega}(t)|=1\pm O(r^2)$ and its Fourier coefficients decay faster
than $r^{|2m|} |m|^{-|m|/2}$. Also, the sum in (\ref{eq:intform}) does
not exceed $O(r^2 t^{-3/2})$ for large $t$, and the $h_m$ decrease with
$m$ faster than $r^{|m|}$.

({\bf 2}) By (\ref{eq:intform}), for times of order $1/\Gamma$ where
$\Gamma=2\Re(\gamma)$, the survival probability for $\omega$ not close
to a resonance decays as $\exp(-\Gamma t)$. 
It follows from our analysis that for small $r$ the final asymptotic
behavior for $t\rightarrow\infty$, is $|\theta(t)|^2=O(t^{-3})$ with
many oscillations as described by (\ref{eq:intform}).

({\bf 3}) When $r$ is larger the polynomial-oscillatory behavior starts
sooner. Since for small $r$ the amplitude of the late asymptotic terms
is $O(r^2)$, increased $r$ yields higher late time survival
probability. This phenomenon, sometimes referred to as atomic
stabilization, can be associated with the perturbation-induced
probability of back-transitions to the well.

({\bf 4}) Using a continued fraction representation 
$\Gamma$ can be calculated convergently for any $\omega$ and $r$. 
The limiting behavior for small $r$ of the exponent $\Gamma$ is
described as follows.  Let $n$ be the integer part of $\omega^{-1}+1$ and
assume $\omega^{-1}\notin\NN$. Then we have, for $T>0$ ($t=r^{-2n}T$),

\begin{equation}\label{limit}\hat{\Gamma}=-T^{-1}\lim_{r\rightarrow 0}
\ln\left|\theta (r^{-2n}T)\right|^2
=\frac{2^{-2n+2}\sqrt{n\omega-1}}{\displaystyle{n\omega}\prod_{m<
n}(1-\sqrt{1-m\omega})^2}
\end{equation}

({\bf 5}) The behavior of $\Gamma$ is different at the resonances
$\omega^{-1}\in\NN$. For instance, whereas if $\omega$ is not close to
$1$, the scaling of $\Gamma$ implied by (\ref{limit}) is $r^2$ when
$\omega>1$ and $r^4$ when $\frac{1}{2}<\omega<1$, by taking
$\omega-1=r^2/\sqrt{2}$ we find 

\begin{equation}
  \label{reson}
  -T^{-1}\lim_{\begin{subarray}{c} r\rightarrow 0 \\ \omega=1+r^2/\sqrt{2}
\end{subarray}}
\ln\left|\theta (r^{-3}T)\right|^2
=\frac{2^{1/4}}{8}-\frac{2^{3/4}}{16}
\end{equation}

\subsection{General periodic case}
We write
\renewcommand{\labelenumi}{(\alph{enumi})}

\begin{equation}
  \label{defeta}
  \eta=\sum_{j=0}^{\infty}\Big(C_j e^{i\omega j t}+C_{-j}e^{-i\omega j
t}\Big)
\end{equation}

\z Our assumptions on the $C_j$ are \  \ \ (a) $0\not\equiv\eta\in L^{\infty}(\mathbb{T})$, (b) $C_0=0$ and (c)
$C_{-j}=\overline{C_j}$.

\z {\bf Genericity condition (g)}.\hfill\break Consider the right shift operator ${T}$
on $l_2(\NN)$ given by ${T}(C_1,C_2,...,C_n,...)$ $=(C_2,C_3,...,C_{n+1},...)$.
We say that ${\bf  C}\in l_2(\NN)$ is {\em generic with respect to
${T}$}
if the Hilbert space generated by all the translates of ${\bf C}$ contains
the vector $e_1=(1,0,0...,)$ (which is the kernel of ${T}$):

\begin{equation}
  \label{eq:cyclic}
e_1\in\bigvee_{n=0}^{\infty}{T}^n \bf C
\end{equation}

\z (where the right side of (\ref{eq:cyclic}) denotes the closure of the
space generated by the ${T}^n \bf C$ with $n\geq 0$.) This condition is
generically satisfied, and is obviously weaker than the ``cyclicity''
condition $ l_2(\NN)\ominus \bigvee_{n=0}^{\infty}{T}^n \bf C=\{0\}$,
which is also generic.

An important case, which satisfies (\ref{eq:cyclic}), (but fails the
cyclicity condition) corresponds to $\eta$ being a trigonometric
polynomial, namely $\bf C\not\equiv 0$ but $C_n=0$ for all large enough
$n$. 

 A simple example which fails
(\ref{eq:cyclic}) is

\begin{equation}
  \label{eq:counterex}
  \eta(t)=2r\lambda\frac{\lambda-\cos(\omega
    t)}{1+\lambda^2-2\lambda\cos(\omega t)}
\end{equation}
\z for some $\lambda\in (0,1)$, in which case $C_n=-r\lambda^n$ for $n\geq
1$. In this case the space generated by $T^n\bf C$ is
one-dimensional. We  prove that there are values of $r$ and
$\lambda$ for which the ionization is incomplete, i.e.\ $\theta(t)$ does
not go to zero for large $t$.

\subsection{Results in the periodic case}

\begin{Theorem}[\cite{[21]}]\label{T11}
  Under assumptions (a), (b), (c) and (g), the survival probability $P(t)$ of
the bound state $u_b$, 
  $|\theta(t)|^2$ tends to zero as $t\rightarrow\infty$.
\end{Theorem}

\begin{Theorem}[\cite{[21]}]\label{T2}
  For $\psi_0(x)=u_b(x)$  there exist values of
  $\lambda$, $\omega$ and $r$ in (\ref{eq:counterex}), for which
  $|\theta(t)|\not\rightarrow 0$ as
  $t\rightarrow\infty$.
\end{Theorem}

\z {\bf Remarks}.

1. Theorem~\ref{T11} can be extended to show that $\int_D|\psi(x,t)|^2dx
\rightarrow 0$ for any compact interval $D\in\RR$. This means that the
initially localized particle  wanders off to infinity since by
unitarity of the evolution $\int_\RR|\psi(x,t)|^2dx=1$.
Theorem~\ref{T2} can be extended to show that for some fixed $r$ and
$\omega$ in (\ref{eq:counterex}) there are infinitely many $\lambda$,
accumulating at $1$, for which $\theta(t)\not\rightarrow 0$. In these
cases, it can also be shown that for large $t$, $\theta$ approaches a
quasiperiodic function.

2. There is a direct connection between our results and Floquet theory
where, for a time-periodic Hamiltonian $H(t)$ with period
$T=2\pi/\omega$, one constructs a quasienergy operator (QEO). Complete
ionization thus corresponds to the absence of a discrete spectrum of the
QEO and conversely stabilization implies the existence of such a
discrete spectrum.  In fact, an extension of Theorem~\ref{T2} shows that
for the initial condition $\psi_0 = u_b$, $\psi_t$ approaches such a
function with $\mu = -s_0$.  More details about Floquet theory and
stability can be found in \cite{[30],[31]}

3.  We are currently investigating extensions of our results to the case
where $H_0=-\nabla^2+V_0(x)$, $x\in\RR^d$, has a finite number of bound
states and the perturbation is of the form $\eta(t)V_1(x)$ and both
$V_0$ and $V_1$ have compact support.  Preliminary results indicate
that, with much labor, we shall be able to generalize Theorem~\ref{T11},
to generic $V_1(x)$.  The definition of genericity will, however, depend
strongly on $V_0$.

\end{document}